\documentclass[twocolumn]{openjournal}

\usepackage{orcidlink}
\usepackage[T1]{fontenc}
\usepackage{ae,aecompl}
\usepackage{graphicx}	
\usepackage{amsmath}	
\usepackage{amssymb}	

\usepackage{natbib}

\usepackage{hyperref}
\hypersetup{
	colorlinks=true,
	urlcolor=blue,    
	linkcolor=black,  
	citecolor=blue,   
}

\defcitealias{Pittordis_2018}{PS18}
\defcitealias{Pittordis_2019}{PS19}
\defcitealias{Badry_2021}{ERH}
\defcitealias{Banik_2018_Centauri}{BZ18}

\hypersetup{citecolor=blue, 
            linkcolor=red, 
            menucolor=blue, 
            urlcolor=blue}  

\newcommand{\km}{{~\rm km}}
\newcommand{\s}{{~\rm s}}

\newcommand{\erg}{{~\rm erg}}
\newcommand{\yr}{{~\rm yr}}
\newcommand{\Myr}{{~\rm Myr}}
\newcommand{\Gyr}{{~\rm Gyr}}

\newcommand{\keV}{{~\rm keV}}

\newcommand{\AU}{{~\rm AU}}

\begin{document}

\title{Supernovae in 2023 (review): possible breakthroughs by late observations}


\author{Noam Soker\,\orcidlink{0000-0003-0375-8987}} 
\affiliation{Department of Physics, Technion, Haifa, 3200003, Israel;  soker@physics.technion.ac.il}

\begin{abstract}
I present a review of how late observations of supernovae, of the nebular phase, and much later of supernova remnants (SNRs), and their analysis in 2023 made progress towards possible breakthroughs in supporting the \textit{jittering jets explosion mechanism (JJEM)} for core-collapse supernovae (CCSNe) and in introducing the \textit{ group of lonely white dwarf (WD) scenarios} for type Ia supernovae (SNe Ia). The new analyses of CCSN remnants (CCSNRs) reveal point-symmetric morphologies in a way unnoticed before in several CCSNRs. Qualitative comparison to multipolar planetary nebulae that are shaped by jets suggests that jets exploded these CCSNe, as predicted by the JJEM, but incompatible with the prediction of the delayed neutrino explosion mechanism. The spherical morphology of the ejecta Pa 30 of the historical type Iax supernova (SN Iax) of 1181 AD, which studies in 2023 revealed, is mostly compatible with the explosion of a lonely WD. Namely, at the explosion time, there is only a WD, without any close companion, although the WD was formed via a close binary interaction, i.e., binary merger. Identifying point-symmetry in SNR G1.9+0.3, a normal SN Ia and the youngest SN in the Galaxy, suggests an SN explosion of a lonely WD inside a planetary nebula (an SNIP). The group of lonely WD scenarios includes the core degenerate scenario and the double degenerate scenario with a merger to explosion delay (MED) time. SN Ia explosions of lonely WDs are common and might actually account for most (or even all) normal SNe Ia.     
\end{abstract}

\keywords{Supernovae: general -- ISM: supernova remnants -- stars: massive -- stars: neutron --  stars: white dwarfs --  stars: jets}

\section{Introduction} 
\label{sec:intro}

The objective of this review is to present the topics that supernova (SN) studies cover in 2023 in relation to the origin and explosion mechanisms of SNe. I review both core-collapse SNe (CCSNe; Section \ref{sec:CCSNe}) and type Ia SNe (SNe Ia; Section \ref{sec:SNeIa}). I emphasize some challenges that new studies in 2023 present and how these studies advance our knowledge toward possible breakthroughs.
 
As such, this review focuses on papers from the year 2023 that are relevant to either the explosion mechanism of CCSNe or to the correspondence between different scenarios of SNe Ia and observed types of SNe Ia, including normal SNe Ia and peculiar SNe Ia. I will not review the basic physics and observations of these supernovae, nor will I review pair-instability SNe (e.g., \citealt{Schulzeetal2023}). I will almost not cite papers before the year 2023 as these are cited in the relevant papers from 2023 that I cite. 
For extended reviews of CCSN explosion mechanisms, the reader should consult earlier reviews, e.g., \cite{Jankaetal2016R} and \cite{Mazzacappa2023} for reviews of the physics of CCSNe and the delayed-neutrino explosion mechanism and \cite{Soker2022RAARev} on the jittering jets explosion mechanism (JJEM). In \cite{Soker2022RAARev}, I argued that models that attribute the extra energy of super-luminous CCSNe in the frame of the delayed neutrino explosion mechanism to magnetars must include jets because jets during and/or after the explosion accompany the formation of energetic magnetars. I will return to this important point in Section \ref{subsec:Explosion}. 
 
The different scenarios of SNe Ia motivated many reviews on the topic of SNe Ia in the last decade which together cover all aspects of SNe Ia before 2023 (e.g., \citealt{Maozetal2014, MaedaTerada2016, Hoeflich2017, LivioMazzali2018, Soker2018Rev, Soker2019Rev, WangB2018,  Jhaetal2019NatAs, RuizLapuente2019, Ruiter2020, Liuetal2023Rev}). \cite{Vinkoetal2023} present a short review of some recent aspects of SNe Ia and CCSNe. For observational classification, see, e.g., \cite{Aleoetal2023}.

CCSNe and SNe Ia differ in many aspects from each other. They also have some common physical processes. The common ground of CCSNe and SNe Ia in this review is the usage of supernova remnants (SNRs). I will use the morphologies of some SNRs to reach what I consider to be significant steps towards breakthroughs in 2023. One breakthrough is coming closer to answering the question of \textit{what is the explosion mechanism of CCSNe?} (Section \ref{subsec:Breakthrough}) and the other breakthrough is in establishing the \textit{group of lonely white dwarf (WD) scenarios} of SNe Ia, which I argue hints at the dominant scenarios of normal SN Ia (Section \ref{subsec:LonelyWD}). I start (Section \ref{sec:PointSymemtry}) by describing the point-symmetric morphology that is relevant to the SNRs that I describe.  I open each of the following two Sections, Section \ref{sec:CCSNe} on CCSNe and Section \ref{sec:SNeIa} on SNe Ia, with a review of relevant papers from 2023 and then present new results that I claim might be considered breakthroughs. 
I summarize this review in Section \ref{sec:Summary}.

\section{The point-symmetric morphology of SNRs} 
\label{sec:PointSymemtry}

Because the study of SNR morphology is relevant to CCSNe and to the progenitors of SNe Ia, I discuss the point-symmetry morphology here. 

The study of SNRs (their properties, their progenitors, their interaction with the ISM, particle acceleration, and more) is a subject by itself that continued to be a major topic in 2023 (e.g., \citealt{Araya2023, Bakisetal2023, Balletal2023, Bambaetal2023, Banovetzetal2023, Bozzettoetal2023, Braunetal2023, Bykovetal2023, CaldwellRaymond2023, Caoetal2023, Chamberyetal2023, Dengetal2023, Dokaraetal2023, Domceketal2023, DuvidovichPetriella2023, Eagleetal2023, Ellienetal2023, Ferrazzolietal2023, Fujishigeetal2023, Ghavamianetal2023, Godinaudetal2023, Himonoetal2023, HollandAshfordSlaneLong2023, Katsuda2023, Khabibullinetal2023, KimKooOnaka2023, Koplitzetal2023, Makarenkoetal2023, Meyeretal2023, Naritaetal2023, NguyenDangetal2023, Picquenotetal2023, Picquenotetal2024, RanasingheLeahy2023, RhoRaviSlavinCha2023, RhoRavietal2023, Sankritetal2023, Sanoetal2023, SewardPoints2023, Shanahanetal2023, SharmaPetal2023, SinitsynaSinitsyna2023, Suherlietal2023, SushchBrose2023, Suzukietal2023, VasilievShchekinov2023, WinklerLongBlair2023, Winkleretal2023, Xiaoetal2023, XinYGuo2023, YamauchiPannuti2023, Yeungetal2023, ZhangMLi2023, ZhangQQetal2023, ZhangPXin2023, Zhongetal2023, Zhouetal2023, Alsaberietal2024, AlsaberiFilipovicetal2024, Ariasetal2024, Baoetal2024, Feseonetal2024, Khabibullinetal2024, Kirchschlageretal2024, LopezCaraballoetal2024, Luoetal2024, Mantovaninietal2024, MichailidisPBetal2024, MichailidisPSetal2024, Milisavljevicetal2024, ReyesIturbideetal2024, ReynoldsBorkowski2024, Romanoetal2024, SatoMatsunagaetal2024, Vinketal2024, ZhangZetal2024, ZhouXetal2024}). I will focus only on the morphologies of some SNRs. 

A point symmetric morphology is one where, for (almost) each of the major morphological features in a nebula or a remnant, there is an opposite morphological feature with respect to the center. Such morphological features might be clumps, filaments, lobes, and ears. An ear is a protrusion from the main nebula or remnant that has a base smaller than the main nebula/remnant and a cross-section that decreases outwards. \cite{Cottonetal2023} argue that ears are common in SNRs. This is also evident from the classification of CCSN remnants (CCSNRs; \citealt{Soker2023class}). 

The point-symmetry morphology is the main SNR property that was used in 2023 to point to a possible breakthrough in our understanding of CCSN explosion mechanisms and to establish the group of lonely WD SN Ia scenarios. I use the large body of knowledge that exists in the study of bipolar and point-symmetric planetary nebulae, like multi-polar planetary nebulae. Many studies of planetary nebulae show that bipolar and point-symmetric structures are shaped by jets (e.g., \citealt{Akashi2023, GomezMunozetal2023, Loraetal2023, Marietal2023, MoragaBaezetal2023}  for some 2023 studies). Comparison of some SNR morphologies to point-symmetric planetary nebulae can lead to the identification of point-symmetric SNRs (e.g., \citealt{Soker2023class, Soker2024IAU384}), as is the comparison to point-symmetric X-ray morphologies in groups and clusters of galaxies \citep{Soker2024CFPointSymmetry}. 
  
The interaction of SN ejecta with a circumstellar material (CSM) or with the interstellar medium (ISM) cannot shape a point-symmetry morphology unless the ejecta and/or the CSM have point-symmetric morphology, to begin with; I here mention only studies from 2023 (and early 2024; see discussion in \citealt{SokerG19032024}). 
A magnetized medium cannot form a point-symmetry with three or more pairs of opposite ears/clumps/filaments/lobes as they form a structure with one symmetry axis, or rarely two symmetry axes, one of the exploding jets and one of the ambient magnetic field direction (e.g., \citealt{Meyeretal2024}).  
\cite{Velazquezetal2023} simulated a non-spherical wind interaction with a magnetized ISM. They find that in a pre-explosion axisymmetric wind with an axis inclined to the interstellar medium (ISM) magnetic field, the ears of the SNR might be bent. Namely, not exactly on a straight line through the center. However, this process cannot shape point-symmetric filaments/clumps/ears/lobes. \cite{ZhangSTianetal2023} included magnetic fields and ISM density gradients in their simulation of SNR G1.9+0.3 (that I study in Section \ref{subsubsec:G1903}). They could form an SNR with a pair of ears but not a point-symmetric SNR. \cite{Villagran2023} performed magneto-hydrodynamic simulations aiming at shaping SNR G1.9+0.3. They blew a non-spherical pre-explosion wind into a magnetized ISM. They obtained an axisymmetrical morphology but not a point-symmetric morphology. Instabilities that are excited by ejecta-ISM interaction, as \cite{Mandaletal2023} study, cannot form point-symmetric morphology. Turbulence in the interstellar medium (ISM), e.g., \cite{RigonInoue2023}, also cannot form point-symmetric structures. 

My conclusion is that point-symmetric SNRs are shaped by jets. In this review, I present results from 2023 (including early 2024) attributing the point-symmetric structures of three CCSNRs (N63A, Vela, SN 1987A; for five more point-symmetric CCSNRs see \citealt{Soker2024CFPointSymmetry}) to point-symmetric ejecta shaped by jets that were part of the pairs of jets that exploded the star in the frame of the JJEM, while a point-symmetric SNR Ia (SNR G1.9+0.3) is attributed to a point-symmetric CSM which was a planetary nebula. 

\section{Core collapse supernova ejecta} 
\label{sec:CCSNe}

\subsection{Neutrino-driven versus jet-driven explosions} 
\label{subsec:Explosion}

I start by reviewing some (but not all) studies of CCSNe that are relevant to the quest for the explosion mechanism. In comparing theory with observations, one must bear in mind the large diversity of CCSNe, as studies in 2023 continued to show (e.g., \citealt{Agudoetal2023, Ailawadhietal2023, Aryanetal2023b, BenAmietal2023, Blanchardetal2023, BostroemDessartetal2023, BostroemZapartasetal2023, ChenPGalYam2023, Corsietal2023, Dasetal2023, DesaiAshalletal2023, DongYValentietal2023, DornWallensteinetal2023, Gangopadhyayetal2023, Holmboetal2023, IraniMoragetal2023, JeenaBanerjeeHeger2023, JiangLiu2023, JinYoonBlinnikov2023, KilpatrickIzzoetal2023, KonyvesTothSeli2023, Kuncarayaktietal2023, LiZhongDai2023, LinHetal2023, Matsuokaetal2023, Mooreetal2023, MoriyaSubrayanetal2023, ParkYoonBlinnikov2023, PessiPrietoAnetal2023, PessiPrietoDessart2023, Pursiainenetal2023b, RodriguezMaozNakar2023, Scullyetal2023, Shivkumaretal2023, Sitetal2023, Srinivasaragavanetal2023, StritzingerBaronetal2023, Stritzingeretal2023, Subrayanetal2023, SunNetal2023, TejaSinghSahuetal2023, UtrobinChugai2024, VanDykBostroemetal2023, VasylyeVoglvetal2023, Vazquezetal2023, WangQetal2023, WangQGoetal2023, WenXGaoetal2023, Williamsonetal2023, YanSWangetal2023, ZhuJetal2023, Bjornsson2024, Brennanetal2024b, Burrowsetal2024b, ChugaiUtrobin2024, DasKetal2024, Dastidaretal2024, DessartGutierrezetal2024, Dessartetal2024, EliasRosaetal2024, Kangasetal2024, Linetal2024, Niuetal2024, Satoetal2024, Shahbandehetal2024, SharmaYetal2024}).    
  
I refer here to two explosion mechanisms: the delayed neutrino explosion mechanism and the JJEM.  
There are many new studies in 2023 of different aspects of the delayed neutrino explosion mechanism (e.g., \citealt{AguileraDenaetal2023, Barkeretal2023, Bocciolietal2023, BurrowsVartanyanWang2023, GogilashviliMurphyMiller2023, Navoetal2023, Orlando2023, Powelletal2023, SieverdingKresseJanka2023, Sumietal2023, vanBaaletal2023, WangTBurrows2023a, WangTBurrows2023b, Zhaetal2023, Andresenetal2024, GhodlaEldridge2024, JankaKresse2024, Maunderetal2024, Schneideretal2024, WangTBurrows2024}; for a recent review see \citealt{BoccioliRoberti2024}). \cite{Varmaetal2023}, for example, find with their 3D simulations that magnetic fields help drive earlier and more energetic explosions (see also \citealt{MatsumotoTakiwakiKotake2024}).
There is a lower number of new studies of the JJEM (e.g., \citealt{AntoniQuataert2023, BearSoker2023,  ShishkinSoker2023, Soker2023kick}), including the operation of the JJEM in electron capture supernovae \citep{WangShishKinSoker2024}, that potentially can take place for stars with an initial mass of $8.5 M_\odot \lesssim M_{\rm ZAMS} \lesssim 9.2M_\odot$ (e.g., \citealt{GuoYLetal2024, Limongietal2024}).  I will devote Section \ref{subsec:Breakthrough} to describe new results that I consider to support the JJEM strongly and are toward a breakthrough in answering the question of the CCSN explosion mechanism. I therefore hope this review will motivate more studies of the JJEM. 

A short description of the JJEM is in place here. The JJEM operates even if the pre-collapse core does not rotate. The idea is that instabilities above the newly born NS amplify stochastic angular momentum fluctuations that were formed in the convective layers of the pre-collapse core. The amplified angular momentum fluctuations are sufficiently large, according to the conjecture that needs future confirmation, to form stochastically varying intermittent accretion disks. These disks launch pairs of jets with varying directions and intensities. 
In total, during the explosion process that lasts for $\simeq 1 - 10 \s$ or even longer (e.g., \citealt{WangShishKinSoker2024}), there might be $\approx {\rm few} - 30$ jet-launching episodes, each lasting for $\simeq 0.01-0.1 \sec$ \citep{PapishSoker2014a}. The jets launching velocities are $\simeq 10^5 \km \s^{-1}$ (compatible with most CCSNe not having relativistic jets, e.g. \citealt{Guettaetal2020}). In each jet-launching episode, the jets carry a mass of $\approx 10^{-3} M_\odot$. These values imply that the newly born NS accretes a mass of $\approx 0.1 M_\odot$ through intermittent accretion disks and that each accretion disk of an episode has a mass of $\approx 10^{-2} M_\odot$. 
Regarding the basic outcomes of the explosions, e.g., nucleosynthesis and lightcurves,  for a given pre-explosion stellar model and the explosion energy, the JJEM's expectation is that nucleosynthesis and lightcurves are very similar to those in the neutrino-driven mechanism. It is important to note that the JJEM does include heating by neutrinos but as a boosting process rather than the sole or the main effect  \citep{Soker2022Boosting}.  In the case of energetic CCSNe beyond the reach of the models of the delayed-neutrino explosion mechanism, there is a need for dedicated simulations to calculate the outcomes since, in those cases, the JJEM requires more energetic jets.  The major difference between the JJEM and the neutrino-driven explosion mechanism regarding the SN ejecta is the morphology of the ejecta, which is one of the major topics of this review and a new study (\citealt{Soker2024CFPointSymmetry}).

The JJEM fundamentally differs (e.g., \citealt{Soker2022RAARev}) from the fixed-axis mechanism of the magnetorotational-driven CCSNe (e.g., \citealt{Khokhlovetal1999, LopezCamaraetal2013, BrombergTchekhovskoy2016, Gottliebetal2022, ObergaulingerReichert2023, Urrutiaetal2023a, Muller2024}): (1) According to the JJEM jets explode most, and possibly all, CCSNe, and not only rare CCSNe. (2) The JJEM operates even when the pre-collapse core does not rotate.  (3) According to the JJEM there are no massive stars that end without explosion. Namely, there are no failed CCSNe. Even when a black hole is formed, there is an explosion driven by jets. (4) The JJEM operates in a jet-negative feedback mechanism: when the jets manage to expel the core in the explosion process, the accretion process of collapsing gas stops (with some delay time). This results in cases where the total explosion energy is several times the binding energy of the ejected mass. 

Both of these explosion mechanisms require instabilities in the zone between the newly born neutron star (NS) and the stalled shock of the collapsing core at $r \simeq 100-150 \km$. The delayed neutrino explosion mechanism requires the instabilities to revive the stalled shock (e.g., \citealt{Fryeretal2023GWnu}). The JJEM requires the instabilities to increase the angular momentum seed perturbations that the convective zones of the collapsing core supply, in particular, the spiral modes of the standing accretion shock instability (SASI). For a recent study of the spiral SASI, see \cite{Buelletetal2023}.   

Some CCSNe require a central engine to operate after the explosion itself (e.g., \citealt{Matsumotoetal2024}). Many studies attribute the extra energy to a rapidly rotating magnetized NS, i.e., a magnetar (e.g.,  \citealt{Acharyyaetal2023, Baietal2023, ChenZetal2023a, Gkinietal2023, Huetal2023, OmandSarin2023, Prasannaetal2023, Tinyanontetal2023, WangTWangSetal2023}). As I pointed out in the past, the formation process of an energetic magnetar is accompanied by the launching of even more energetic jets during and/or after the explosion. The majority of magnetar studies continue to ignore the role of jets. The problems of fitting the lightcurves of superluminous CCSNe with magnetars continue with 2023 papers. In \cite{Soker2022RAARev}, I analyzed the paper by \cite{ChenZetal2023b}, who fitted the lightcurves of 70 hydrogen-poor superluminous CCSNe with magnetars; about a third of these have an initial spin period of $P_{\rm mag} \le 0.002 \s$. However, such rapidly rotating magnetars will likely be spun up during the last phase of their formation by an accretion disk/belt that launches jets. I argued that even the slower magnetars in their fitting most likely launched jets at the explosion. In seven out of the 14 CCSNe they fit with ejecta-CSM interaction, the explosion energy is much larger than what the delayed neutrino explosion mechanism can supply, implying a jet-driven explosion. The claim that supernova studies that include energetic magnetars cannot ignore jets continues to hold in 2023. The same arguments hold for the magnetars with $P_{\rm mag} \le 0.002 \s$ that \cite{Westetal2023} and \cite{Poidevinetal2023} find to fit the hydrogen-poor superluminous SN 2020qlb and SN 2021fpl, respectively, for the number of short-period magnetars that \cite{DongXLiuetal2023} argue to power bumpy lightcurves, for those magnetars with $P_{\rm mag} \le 0.002 \s$ that  \cite{Gomezetal2023} and \cite{Hinkleetal2023} fit to hydrogen-poor superluminous CCSNe, and to the magnetar that \cite{FioreBenettietal2024} fit to SN 2019neq.  

The delayed neutrino explosion mechanism cannot account for energetic CCSNe. \cite{Hiramatsuetal2023} find the explosion energy of the type IIn SN 2021qqp to be $E_{\rm exp} \simeq (3-10)\times 10^{51} \erg$, and \cite{SmithNetal2024} estimate $E_{\rm exp} \simeq (5-10)\times 10^{51} \erg$ for type IIn SN 2015da. The delayed neutrino explosion mechanism cannot account for such explosion energies, while the JJEM can naturally explain such explosion energies (for a discussion from 2023, see, e.g., \citealt{ShishkinSoker2023}). \cite{Salmasoetal2023} deduced that the explosion energy of  SN 2020faa is $\simeq 4 \times 10^{51} \erg$, and discussed the possibility that this peculiar SN II was powered mainly by jets. Likewise,  \cite{Nagaoetal2023} find the explosion energy of the type Icn SN 2021ckj to be $\simeq 4 \times 10^{51} \erg$, an energy the neutrino delayed mechanism cannot account for.  \cite{Pumoetal2023} specifically argue that some of their modeled CCSNe with high explosion energies challenge the neutrino-driven explosion mechanism.  
\cite{Schweyeretal2023} estimate the kinetic energy of SN 2019odp, a type Ib CCSN, to be $E_{\rm exp} \simeq 6\times 10^{51} \erg$ and note that it is above what the delayed neutrino explosion mechanism can supply. \cite{Karamehmetogluetal2023} model several SNe Ibc and find their kinetic energy to be $>3 \times 10^{51} \erg$, more than what the delayed neutrino explosion mechanism can account for. Again, the JJEM easily accounts for such explosion energies. \cite{Ergonetal2023} discuss the possibility that jets can improve their fitting of the lightcurve of the type IIb SN 2020acat. 

Another unsolved problem of the delayed neutrino explosion mechanism is that it seems it cannot produce the observed mass of $^{56}$Ni in stripped-envelope CCSNe \citep{SawadaSuwa2023}. \cite{Imashevaetal2023}, on the other hand, argue that with proper treatment of the collapsing core, they can produce more $^{56}$Ni, at least for the model appropriate for SN 1987A.   

Studies in 2023 continue the exploration of gravitational waves from CCSNe and their detectability (e.g., \citealt{Brueletal2023, Bugli2023,  CasallasLagosetal2023, Dalaysetal2023, Hestonetal2023, Hsiehetal2023, Iessetal2023, Jakobusetal2023, Mezzacappaetal2023, Mitraetal2023, Pajkosetal2023, PastorMarcosetal2023, Shibagakietal2023, Szczepanczyketal2023, Vartanyanetal2023, Villegasetal2023, Wolfeetal2023, YuanYFanetal2023, Nunesetal2024, Zhadetdal2024a}), including from two cocoons (lobes) that are inflated by fixed-axis energetic jets related to CCSNe \citep{Gottliebetal2023} and by several jet-inflated pairs of lobes (cocoons) in the JJEM \citep{Soker2023RAAGW}. The detection of neutrinos from CCSNe continued to be a topic of interest in 2023 (e.g., \citealt{Abbasietal2023, Akahoetal2023, Bendahmanetal2023, Dragoetal2023, Fiorilloetal2023, LinZetal2023, MukhopadhyaySen2023, PagliaroliTernes2024, PrasannaThompsonetal2023, VartanyanBurrows2023nu, Ashida2024, Powelletal2024, Saezetal2024}). The detection of gravitational waves or neutrinos is not at a stage that allows the support of one explosion mechanism or the other (for the study of CCSN diagnostics, see, e.g., \citealt{Fryeretal2023GWnu}). Therefore, I will not touch on these topics. I will also not touch the subject of axion instability supernova (e.g., \citealt{Morietal2023}). 
 
The presence of CSM around CCSN progenitors and the interaction of the CCSN ejecta with this CSM were major subjects in the 2023 supernova literature (e.g., \citealt{Akashietal2023, AndrewsJPearsonetal2023, BenAmietal2023, Bietenholzetal2023, Bilinskietal2023, Bruchetal2023,  ChugaiUtrobin2023, DavisKetal2023, DessartJacobsonGalan2023, Dickinsonetal2023, Ercolinoetal2023, Hiramatsuetal2023, Ishiietal2023, KhatamiKasen2023, Kooetal2023, LinWWangetal2023, MaedaChandraetal2023, MaedaMichiyamaetal2023, MatsuokaSawada2023, Mauerhanetal2023, Moranetal2023, Moriya2023, Pearsonetal2023, Pellegrinoetal2023, Pessietal2023, Petruketal2023, Pursiainenetal2023a, Salmasoetal2023, Sarmahetal2023, Shresthaetal2023, Takeietal2023, TsunaMuraseMoriya2023, TsunaTakei2023, Brennanetal2024a, Dasetal2024, Ferrarietal2024, JacobsonGalanetal2024, Murase2024, MezaRetamaletal2024, Muraietal2024, Sfaradietal2024}), including the role of binary interaction in expelling pre-explosion CSM (e.g., \citealt{Zenatietal2024}).\footnote{\cite{Soker2013PEO} suggested (based on earlier binary models of similar types of transients) that a sudden pre-CCSN swelling may trigger asymmetric mass loss in binary systems. Five months later, \cite{SmithArnett2014} repeated this suggestion. Later papers (e.g., \citealt{Smith2017, Smithetal23ixf2023, Bilinskietal2023, Mauerhanetal2023, SmithNetal2024}) wrongly and consistently have been attributing this suggestion to \cite{SmithArnett2014}.}
The ejecta-CSM interaction has become a major topic in relation to the type II SN~2023ixf (e.g., \citealt{Berstenetal2023, Chandraetal23ixf2023, Guettaetal23ixf2023, Hiramatsuetal23ixf2023, Hosseinzadehetal23ixf2023, Koenig2023, Liueal23ixf2023, Martinezetal2023, Niuetal23ixf2023, PledgerShara2023, QinYetal2023, Sarmah2023,  Sgroetal2023, Tejaetal23ixf2023, VanDyketal23ixf2023, Vasylyevetal23ixf2023, XiangMoetal2023, Yamanakaetal2023, Zimmermanetal2023}). 

Studies found that the ejecta of SN~2023ixf interacted with a CSM that was extended up to a radius of $R_{\rm CSM} \simeq 20-50 \AU$ around SN~2023ixf (e.g., \citealt{Bergeretal2023,  Bostroemetal2023, Grefenstetteetal2023, JacobsonGalanetal2023, Kilpatricketal2023, Tejaetal23ixf2023, Smithetal23ixf2023, ZhangLinWangetal2023, LiHuLietal2024}). In principle, a close CSM might result from an enhanced mass loss rate shortly before the explosion, most likely accompanied by an outburst (eruption; e.g., \citealt{Tsunaetal2023, Tsunaetal2024}) that might be triggered by activity in the core (e.g. \citealt{CohenSoker2024}). However, several studies find no evidence for pre-explosion outbursts in the years before the explosion of SN 2023ixf (e.g.,  \citealt{Dongetal2023, Flinneretal2023, Neustadtetal2023, Jencsonetal2023, Soraisametal2023, Ransomeetal2024}). This challenges any model where the massive CSM is attributed to impulsive mass ejection years before the explosion of SN~2023ixf. The progenitor of SN~2023ixf did have large amplitude pulsations (e.g., \citealt{Kilpatricketal2023, Soraisametal2023}). These pre-explosion properties led to the suggestion that instead of a pre-explosion impulsive ejection of a massive CSM, the progenitor of SN~2023ixf maintained an extended zone above its photosphere of rising and falling dense clumps, i.e., the effervescent zone model \citep{Soker23ixf2023}. 

\cite{ObergaulingerReichert2023} study nucleosynthesis in fixed-axis jets that power some energetic CCSNe, and find that they have enhanced synthesis of iron-group elements. \cite{Reichertetal2024} find that when magnetic fields are included in the fixed-axis simulations, nucleosynthesis can yield up to the third peak of the r-process (but see \citealt{Zhaetal2024b}). \cite{Leungetal2023} and \cite{LeungNomoto2024} study nucleosynthesis in CCSNe driven by fixed-axis jets, including collapsar where the remnant is a black hole.  \cite{Wanajo2023} studies nucleosynthesis in neutrino-heated ejecta and neutrino-driven winds (see also \citealt{Prasannaetal2024}), and \cite{Friedlandetal2024} and \cite{Psaltisetal2024} study the $\nu p$-process in neutrino-driven outflows in CCSNe. These papers do not refer to jets.  
In general, r-process nucleosynthesis in CCSNe was not a major topic in 2023. \cite{Anandetal2023}, for example, study the lightcurves of broad-line stripped-envelope CCSNe and find no evidence for r-process nucleosynthesis. \cite{Pateletal2024} study the effect of r-process nucleosynthesis on the lightcurve of hydrogen-rich CCSNe. 
Nucleosynthesis can be used to study CCSNe. \cite{Hoppe2023} and \cite{LiuNetal2024} present interesting studies of using solar-system isotopes ratio to constrain the stellar properties at the explosion.  

Dust formation in CCSN ejecta is another area of intensive research (e.g., \citealt{NiculescuDuvazetal2023}), as CCSNe might be a significant source of dust in the Universe (e.g., \citealt{Shahbandehetal2023}).

Several papers in 2023 (e.g., \citealt{Reichertetal2023, Volpatoetal2024}) study the rare group of `magnetorotational supernovae', which are driven by fixed-axis jets due to rapid pre-collapse core rotation. While in the delayed neutrino explosion mechanism, these rare CCSNe form a separate group (e.g., \citealt{FujibayashiSekiguchietal2023, MaedaSuzukiIzzo2023}), in the JJEM, they are just the rapidly rotating tail of CCSNe with slow pre-collapse core rotation (e.g., \citealt{Soker2023gap}).
In the fixed-axis case, the rapid pre-collapse rotation of the core forces the accretion disk around the newly born NS (and in some cases later around the black hole) to maintain a constant axis, and so are the jets that the accretion disk launches. This leads to a bipolar explosion (e.g., \citealt{FangMaedaetal2023}). 
\cite{VasylyeYangvetal2024} measure high continuum linear polarization in the type IIP supernova 2021yja and discuss the possibilities that it results from CSM interaction or fixed-axis jet explosion. 
Polarization measurements can teach much about the explosion mechanism (e.g., \citealt{Maund2024}). \cite{Nagaoetal2023Polar} conclude their spectropolarimetry study of SNe II by stating that ``the emergence of a global aspherical structure, e.g., a jet-like structure, might be the key ingredient in the explosion mechanism to produce an energetic SN''. This is compatible with the expectation of the JJEM. 

However, such a rapid rotation is rare. On the other extreme, the core rotation is zero, and the angular momentum of the intermittent disks that launch the jets is stochastic, such that the jets are fully jittering. The stochastic angular momentum direction results from seed angular momentum perturbations by the pre-collapse core convection that are amplified by instabilities above the newly born NS.  In the JJEM, there is a continuum from CCSNe driven by fully jittering jets, which result from progenitors with very slowly rotating pre-collapse cores to CCSNe driven by fixed-axis jets that result from rapidly rotating pre-collapse cores. In the frame of the JJEM, this behavior can explain the mass gap between NSs and black holes \citep{Soker2023gap}. 

\subsection{The breakthrough: evidence for exploding jets} 
\label{subsec:Breakthrough}

From Section \ref{subsec:Explosion}, it is evident that although the delayed neutrino explosion mechanism is more popular in the literature than the JJEM, it nonetheless encounters some challenges. Here, I will argue that the point-symmetric morphologies of three CCSNRs that were studied in 2023 strongly support the JJEM and pose severe problems to the neutrino-driven explosion mechanism (for more point-symmetric CCSNe that were studied later in 2024, see \citealt{Soker2024CFPointSymmetry}). 

Late observations, from months post-explosion before the nebular phase (e.g., \citealt{HayniePiro2023, Medleretal2023}), through the nebular phase (e.g., \citealt{DerKacyetal2023, Dessartetal2023,  DessartHillieretal2023,  Ertinietal2023, FangMaeda2023, Liljegrenetal2023, Marguttietal2023, OmandJerkstrand2023, Ravietal2023, RizzoSmithetal2023, Wessonetal2023, Xiangetal2023}, and then the remnant phase (e.g., \citealt{SatoYoshidaetal2023, Zsirosetal2023} and more below) can teach us about the progenitor properties and its environment at the explosion time and on the explosion mechanism.  Before I describe evidence for jet-driven CCSNe by analyzing the CCSNR phase, I refer to the prediction of the delayed neutrino explosion mechanism that some massive stars collapse into a black hole with no explosion (or with only a very faint explosion). This collapse is termed `failed supernovae'.

Some papers in 2023 continue to study this concept of a `failed supernova' (e.g., \citealt{Coughlin2023, daSilvaSchneideretal2023,  DisbergNelemans2023, KurodaShibata2023, LiLZhuetal2023L, Paradisoetal2023}). A `failed supernova' is a prediction of the delayed neutrino explosion mechanism where neutrino heating cannot explode the star which in turn collapses to a black hole. 
On the other hand, in the JJEM, there are no failed CCSNe.
Even if a black hole is formed, there is an explosion, and more likely, an energetic explosion (see review \citealt{Soker2022RAARev}). In the JJEM, a black hole is likely to be formed when the pre-collapse core is rapidly rotating because this leads to fixed-axis jets that are less efficient in exploding the core (\citealt{Soker2023gap}), i.e., an inefficient jet feedback mechanism (for a recent paper on the accretion disk around the central black hole in CCSNe see \citealt{Fujibayashietal2023}). \cite{Soker2023gap} studies the continuum of explosion outcomes from zero pre-collapse core rotation to rapid pre-collapse rotation in the frame of the JJEM.   

The new study by \cite{StrotjohannOfekGalYam2023} and early results of \cite{ByrneFraser2022} question the existence of a large population of `failed supernovae'. The detection of a black hole at the center of an SNR, as tentatively claimed by \cite{Balakrishnanetal2023} for SNR G351.9-0.9, shows that black holes can form in CCSNe, rather than in `failed supernovae'. 
The detection by \cite{Beasoretal2023} of a luminous infrared source at the position of the `failed supernova' candidate N6946-BH1 shows, as they claim, that it better fits a type II intermediate luminosity optical transient (type II ILOT) than a `failed supernova' (but see \citealt{Kochanek2023, KochanekNeustadtStanek2023}). The above-cited studies challenge the prediction of the delayed neutrino explosion mechanism of `failed supernovae'. I reiterate my claim that studies of CCSNe that form black holes cannot ignore jets. 

Let me then turn to a prediction of the JJEM that in some (but not all) cases, the jittering jets imprint a point-symmetric morphology. Most of the jittering jets during the explosion process, typically lasting for about one to several seconds and more, deposit their entire energy in the core of the star and explode it. From that point on, the explosion is similar (but not identical) to that of the delayed-neutrino explosion mechanism. However, in some cases, the few last pairs of jets that are launched after the inner parts of the stars are exploded might leave an imprint on the outer ejecta. In most cases, only the last pair leaves an imprint, forming two opposite `ears' (e.g., \citealt{Soker2023class}). An ear is a protrusion from the main ejecta with a decreasing cross-section with distance from the center. In a minority of cases, more than one pair of jets might leave imprints, forming a point-symmetric morphology. 

In 2023, I identified point-symmetric morphologies in three CCSNRs. In Section \ref{sec:intro} I discussed the point-symmetric morphology and why jets are most likely to shape such morphologies.

\begin{figure}
\begin{center}
\includegraphics[trim=2.6cm 13.1cm 0.0cm 3.0cm ,clip, scale=0.70]{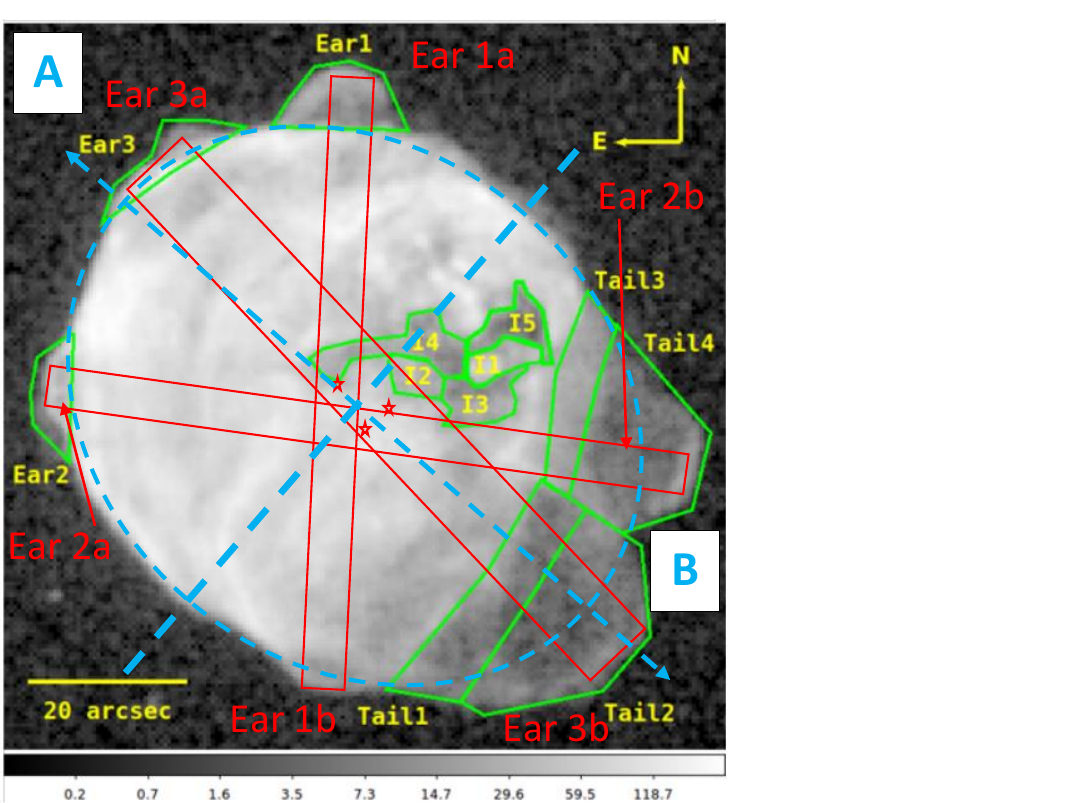}
\end{center}
\caption{The CCSNR N63A has a point-symmetric morphology. 
A Chandra X-ray image of SNR N63A (logarithmic-scale $0.3 -7.0 \keV$). The green lines and the small-yellow writings are from the original figure of \cite{Karagozetal2023}. The red-solid lines, the dashed-blue lines, and the red and blue writing are additions from \cite{SokerN63A2023} that identify a point-symmetric morphology.
}
\label{Fig:N63A} 
\end{figure}

In \cite{SokerN63A2023}, I studied the morphology of the CCSNR N63A based on new X-ray images that \cite{Karagozetal2023} present. 
I present the image and my identification of a point-symmetric morphology in Figure \ref{Fig:N63A}. I identified the point-symmetric morphology to be the three pairs of ears that are marked in the figure. This point symmetry cannot be shaped by ejecta-CSM interaction or ejecta-ISM interaction (Section \ref{sec:intro}). Therefore, this point-symmetry strongly supports the JJEM. Moreover, the differences in sizes of the two ears in each pair are accounted for by the JJEM.   

In each pair, one ear is significantly larger than the opposite ear. This is not due to an asymmetrical medium on sides A and B (marked in the figure), as Ear 1a is larger than Ear 1b, but Ears 2a and 3a are smaller than Ears 2b and 3b, respectively. I, therefore, concluded \citep{SokerN63A2023} that the asymmetry of each pair results from asymmetrical opposite jets at the time of the jets' launching. This asymmetry likely results from the accretion disk that launches the jets in the JJEM having no time to relax fully. In the JJEM there are $\approx {\rm few}$ to $\approx 30$ jet-launching episodes, each lasting for $\approx 0.01-0.1 \s$. The viscous timescale of the accretion disk in each jet-launching episode is not much shorter than this timescale. Therefore, the disk has no time to fully relax to a thin accretion disk. Since the instabilities that must operate in the JJEM are likely to form each accretion disk with unequal sides of the temporarily equatorial plane, this asymmetry lasts for most of the jet-launching episode. The two unequal sides of the accretion disk are very likely to launch two opposite jets that have different energy, mass, and velocity. These asymmetrical two opposite jets in a pair will inflate unequal ears. 
 
\begin{figure}[t]
	\centering
\includegraphics[trim=0.8cm 11.2cm 3.0cm 2.2cm ,clip, scale=0.55]{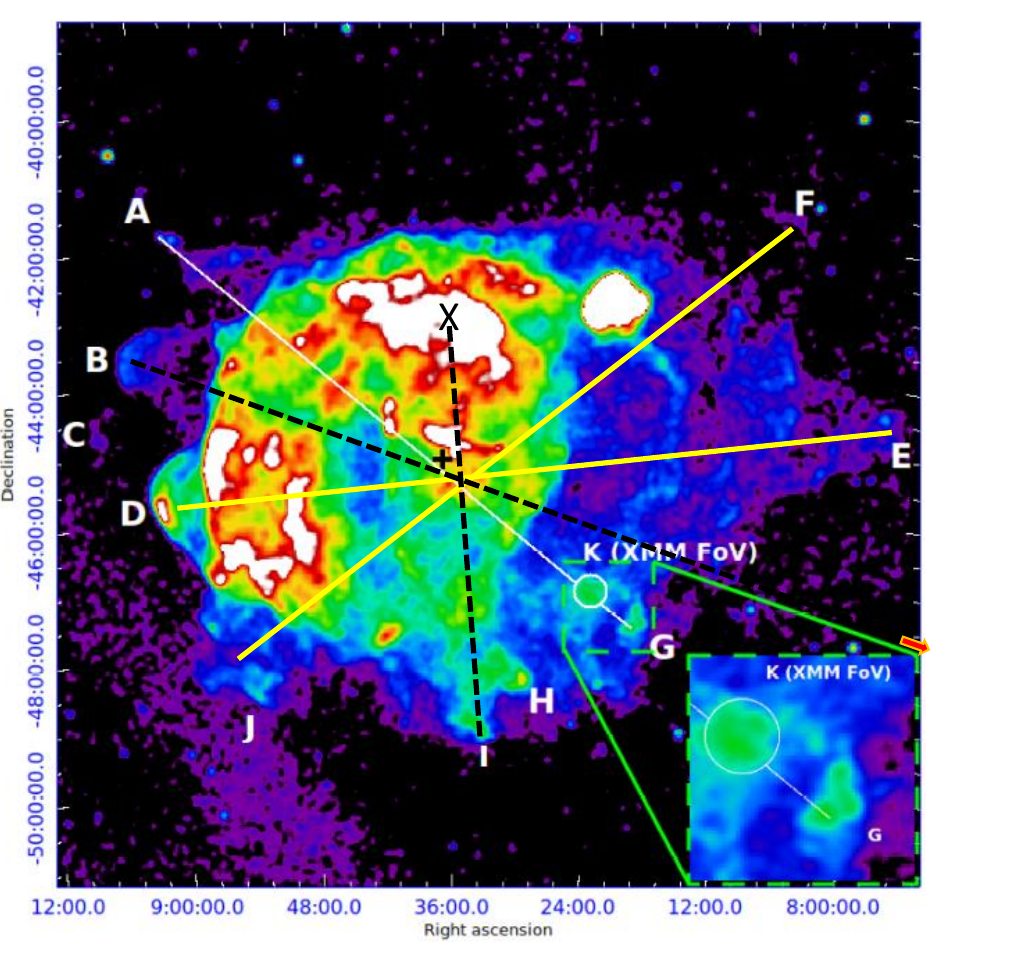} 
\caption{The point-symmetric morphology of the Vela SNR, as identified by \cite{Soker2023class}. 
The ROSAT X-ray image is from \cite{Aschenbachetal1995}, and the figure is adapted from figure 1 of \cite{Sapienzaetal2021}. Clumps A to F were identified by \citealt{Aschenbachetal1995}. The white AG-line and the labeling of the clumps are by \cite{Sapienzaetal2021}. The thick-yellow FJ-line and DE-line and the two dashed-black lines are marked by \cite{Soker2023class}.
Each of the two dashed-black lines connects a clump to an assumed counter jet (for newer analysts, see \citealt{Soker2024CFPointSymmetry}). 
}
\label{Fig:Vela1}
\end{figure}

In \cite{Soker2023class}, I classified 14 CCSNRs into five groups according to their morphology shaped by jets. I argued that the degree of pre-collapse core rotation is one of the main parameters determining the morphological class. The faster the pre-collapse core rotation is the less random the directions of the jets in the different jet-launching episodes are. For a very slow (or zero) pre-collapse core rotation, the jets jitter almost randomly. This, I suggested, might lead to a point-symmetric morphology with pairs of opposite structures in many directions. In that study, I identified a point-symmetric morphology in the Vela SNR.  The Vela SNR is well-studied (e.g., \citealt{Mayeretal2023, Ritchey2023, Kargaltsevetal2024}). I present the identification of the point-symmetric morphology in Figure \ref{Fig:Vela1} taken from \cite{Soker2023class}. Earlier studies already noted the existence of pairs of opposite structural features, as marked in the figure. However, only in \cite{Soker2023class} did I explore the full point symmetry and connect it to jittering jets that exploded the progenitor of the Vela SNR (for newer images of Vela, see \citealt{Mayeretal2023}, and for newer analysis of its point symmetry, see \citealt{Soker2024CFPointSymmetry}).  

SN 1987A has been one of the prime focuses of the SN community since its explosion (for 2023 papers see, e.g., \citealt{Arendtetal2023, Cikotaetal2023, Dohietal2023, Fiorilloetal2023, JonesOetal2023, Kangasetal2023, Larssonetal87A2023, Weishietal2023, Bouchetetal2024, Chenetal2024, Franssonetal2024, Ravietal2024, Sapienzaetal2024}). I am interested in the ejecta morphology and its implications on the explosion mechanism. \cite{Onoetal1987A2023} consider a bipolar explosion morphology for SN 1987A. They do not refer to jets, though.

\begin{figure*}[th]
	\centering
\includegraphics[trim=2.8cm 12.1cm 3.0cm 2.4cm ,clip, scale=0.52]{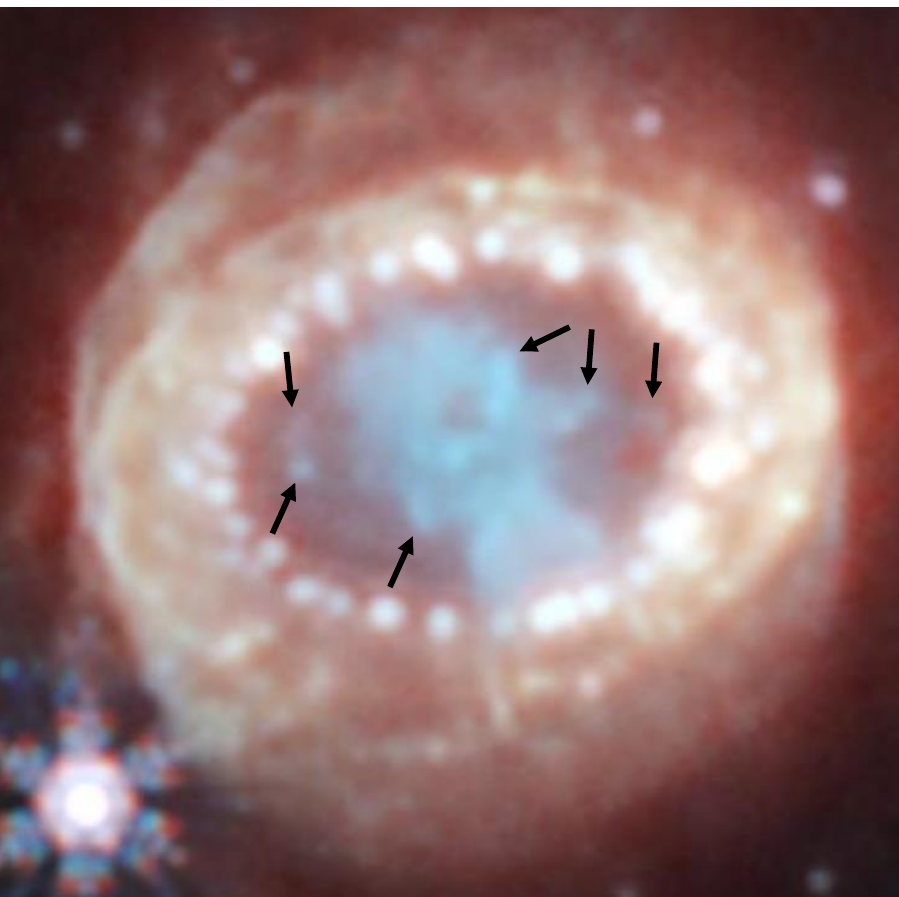} 
\includegraphics[trim=2.8cm 12.1cm 3.0cm 2.4cm ,clip, scale=0.52]{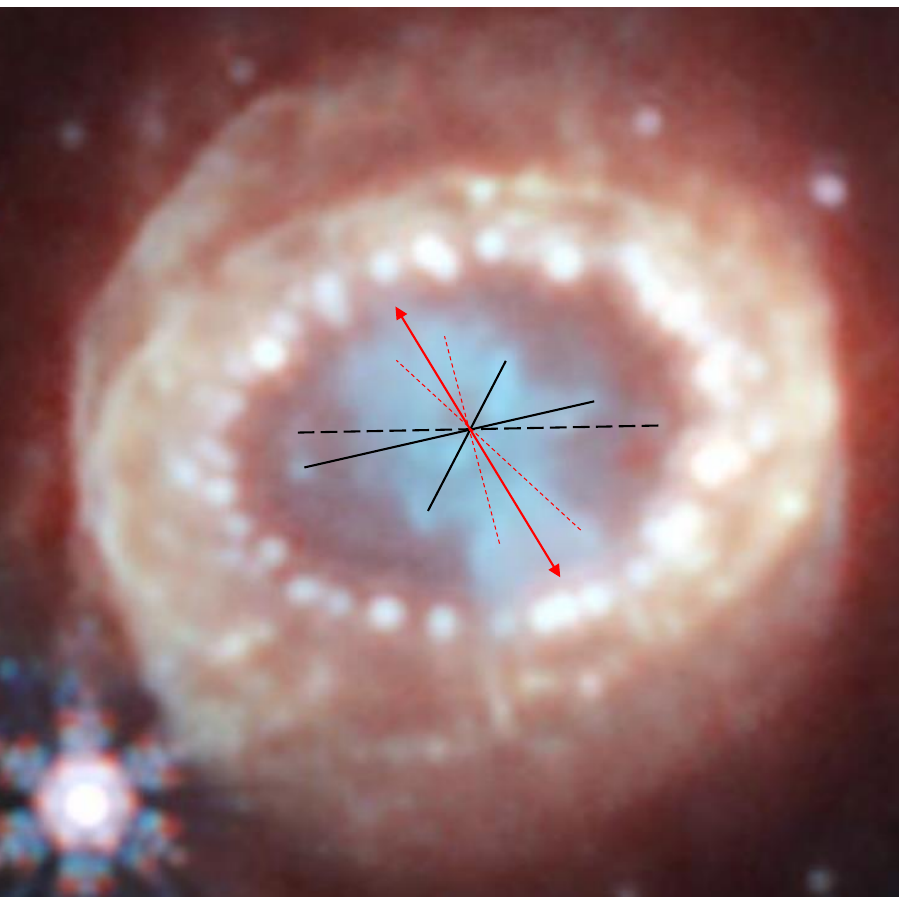} 
\caption{An image of SN 1987A from JWST [source: NASA Webb Telescope Team, AUG 31, 2023; NASA, ESA, CSA: Matsuura, M. (Cardiff University), Arendt, R. (NASA’s Goddard Spaceflight Center \& University of Maryland, Baltimore County), Fransson, C.].\footnote{For the original image see: https://www.nasa.gov/missions/webb/webb-reveals-new-structures-within-iconic-supernova/} 
Red represents emission at 4.44 microns, orange at 4.05 microns, yellow at 3.23 microns, cyan at 1.64 and 2.0 microns, and blue at 1.5 microns. 
The arrows in the left panel are additions from \cite{Soker2024SN1987A}. These arrows point at clumps, namely, some small areas in the outer regions of the blue-cyan structure that are brighter than their surroundings. The three black lines in the right panel, also from \cite{Soker2024SN1987A}, connect pairs of clumps that the arrows in the left panel point at. The red double-headed arrow indicates the direction of the elongated axis of the blue-cyan structure. The two red-dashed lines signify a possible structure of two opposite jets. (for newer images of SN 1987A, see \citealt{Rosuetal2024}, and for newer analysis of its point symmetry, see \citealt{Soker2024CFPointSymmetry}). 
}
\label{Fig:SN1987A}
\end{figure*}

What allows me to (tentatively) claim for the JJEM for 1987A is the new observation of SN 1987A by JWST (\citealt{Arendtetal2023}) as I present in Figure \ref{Fig:SN1987A}. All marks on the original figure are from \cite{Soker2024SN1987A}, where more details can be found. In that paper, I identified a tentative point-symmetric morphology in the new JWST image of SN1987A. As said, the JJEM predicts that in some CCSNe, the last jets imprint a point-symmetrical structure. I suggest such a structure in the ejecta of SN 1987A and suggest that it was exploded by jittering jets. The asymmetry within each pair of two clumps is again attributed to the not relaxed intermittent accretion disks that launched the jets during the explosion. 

Let me end with the study of the northeast jet of the SNR Cassiopeia A by \cite{KooLeeetal2023}. They conclude that the neutrino-driven explosion mechanism and the magnetocentrifugal jet-induced explosion (fixed-axis jets) have problems accounting for the properties of the northeast jet of Cassiopeia A. This is a known result. As I commented in the past, the JJEM can best explain the properties of Cassiopeia A  (e.g., \citealt{Soker2017RAA}). New results (after this paper's first and second versions were posted on the arXiv) by \cite{BearSoker2024} show that Cassiopeia A also has a very robust point symmetry, as expected by the JJEM.
An even later paper \citep{Soker2024CFPointSymmetry} identifies point symmetry in more CCSNe, bringing the number to a total of eight CCSNe with point-symmetric morphology.  That paper presents the images of these eight CCSNRs and compares them to the morphologies of the X-ray emitting gas in some groups and clusters of galaxies that are known to be shaped by jets. This comparison strongly suggests that jets shaped these CCSNRs as well. These point-symmetric CCSNRs are the CCSNR 0540-69.3, which has a point symmetry in its velocity map in a plane along the line of sight and a jet axis on the plane of the sky, and CCSNRs with point-symmetric morphologies on the plane of the sky, Cassiopeia A, N63A, Vela, SN 1987A, G321.3–3.9  which show clear point-symmetry, and   CTB1 and G107.7-5.1 which show less prominent point-symmetry (tentative identification).  

I consider the point-symmetric morphologies of eight CCSNRs, supported by earlier studies of CCSNR morphologies, to be a step toward a breakthrough in determining the CCSN explosion mechanism.  A caveat is that there is not yet a quantitative classification of the point-symmetric morphology of CCSNRs.  The JJEM cannot be ignored when analyzing CCSNe.

\section{Type Ia supernova ejecta} 
\label{sec:SNeIa}

\subsection{SN Ia scenarios} 
\label{subsec:SNIaScenarios}

The diversity of normal and peculiar SNe Ia (e.g., \citealt{DerKacyPaughetal2023, Desaietal2023, Dwarkadas2023RNAAS, Fausnaughetal2023, Grauretal2023, Harrisetal2023, Harveyetal2023, Hinseetal2023, HollandAshfordetal2023, Huangetal2023, Keegansetal2023, Larisonetal2023, LiYZhengetal2023, LuJetal2023, Nascimentoetal2023, OBrienetal2023, OgawaMaedaKawabata2023, PearsonSandetal2023, Petersonetal2023,  SharonKushnir2023, Singhetal2023, Tiwarietal2023, Toyetal2023, XiGWwangetal2023, YangZhangetal2023, Bietal2024, Giuffridaetal2024, Graur2024, Pakmoretal2024, Petreccaetal2024, Singhetal2024, Wengetal2024, Williamsetal2024}) have motivated the consideration of all major SN Ia scenarios in 2023. I classify six scenarios as in \cite{SokerG19032024} and present them in Table \ref{Tab:Table1} (for a different partition of the scenarios, in particular sub-classification of the SD scenario, see the table in the review by \citealt{Liuetal2023Rev}).  
\begin{table*}
\scriptsize
\begin{center}
  \caption{SN Ia scenarios and their ability to form point-symmetric or spherical SNRs}
    \begin{tabular}{| p{1.8cm} | p{2.4cm} | p{2.4cm}| p{2.0cm}| p{2.0cm} | p{2.0cm} | p{2.0cm} |}
\hline  
\textbf{Group} & \multicolumn{2}{c|}{$N_{\rm exp}=1$: \textcolor{red}{\textbf{Lonely WD}}}  &  \multicolumn{4}{c|}{$N_{\rm exp}=2$}     \\  
\hline  
\textbf{Outcome} & \multicolumn{2}{c|}{\textcolor{red}{\textbf{Mostly normal SNe Ia}}}  &  \multicolumn{4}{c|}{\textcolor{red}{\textbf{Mostly peculiar SNe Ia}}}\\ 
\hline  
\textbf{{SN Ia Scenario}}  & {Core Degenerate }    & {Double Degenerate - MED} & {Double Degenerate} & {Double Detonation} & {Single Degenerate} & {WD-WD collision} \\
\hline  
\textbf{{Name}} & CD & DD-MED & DD & DDet & SD-MED or SD & WWC\\
\hline  
\textbf{MED time} & MED  & MED  & 0  &0  & MED or 0 & 0 \\
\hline  
 {$\mathbf{[N_{\rm sur}, M, Ej]}$$^{[{\rm 2}]}$}
  & $[0,M_{\rm Ch},{\rm S}]$ 
  & $[0,M_{\rm Ch}, {\rm S}]$ 
  & $[0,$sub-$M_{\rm Ch},{\rm N}]$
  & $[1,$sub-$M_{\rm Ch},{\rm N}]$
  & $[1,M_{\rm Ch},{\rm S~or~N}]$  
  & $[0,$sub-$M_{\rm Ch},{\rm N}]$ \\
\hline  
\textbf{Point symmetry in the SNR} 
& Expected in SNIPs with point-symmetric PNe. \colorbox{yellow}{{\textcolor{red}{\textbf{SN~1181; G1.9+0.3}}}}
& Very rare: A SNIP with point-symmetric PN.  \colorbox{yellow}{\textcolor{red}{\textbf{SN~1181}}}   
& Very rare: A SNIP with point-symmetric PN.
& Extremely rare. 
& Possible: a symbiotic progenitor; Low-mass CSM. 
& Extremely rare; large-scale departures from ellipticity.  \\              
\hline  
     \end{tabular}
  \label{Tab:Table1}\\
\end{center}

\begin{flushleft}
\small 
Notes: (1) Cells in bold red indicate new suggestions/speculations from 2023 that I support in this review. The two SNRs Ia that I focus on are the peculiar SN 1181, which I suggest might result from the DD-MED or the CD scenarios, and SNR G1.9+0.3, which results from the CD scenario in a planetary nebula (SNIP). 
(2) Abbreviation. MED time: Merger to explosion delay time (includes mass transfer to explosion delay time).    
$N_{\rm exp}$ is the number of stars in the system at the time of the explosion. $N_{\rm sur}=1$ if a companion survives the explosion while $N_{\rm sur}=0$ if no companion survives the explosion; in some peculiar SNe Ia, the exploding WD is not destroyed, and it also leaves a remnant, i.e., $N_{\rm sur}=2$. $M_{\rm Ch}$ for near-Chandrasekhar-mass explosion; sub-$M_{\rm Ch}$ for sub-Chandrasekhar mass explosion. 
The ejecta morphology Ej: S indicates scenarios that can lead to a spherical SNR, while N indicates scenarios that cannot form a spherical SNR.    
\end{flushleft}
\end{table*}
   
In the core degenerate (CD) scenario, a WD merges with the core of an asymptotic giant branch star to form a massive WD. The merger product explodes later as a Chandrasekhar mass ($M_{\rm Ch}$) explosion. This is the merger to explosion delay (MED) time. At the time of the explosion, there is only one relevant star in the system (wide companions do not play any role). Namely, $N_{\rm exp}=1$. The merger process of the CD scenario leaves a planetary nebula descendant. The planetary nebula shell might stay intact for up to $\approx 10^6 \yr$. If the MED time is $\la 10^6 \yr$, the explosion might occur inside the shell, i.e., an SN inside a planetary nebula (SNIP). 
Some peculiar SNe Ia might also be SNIPs, e.g., as I speculated for SNR G352.7-0.1 \citep{Soker2024G352}. 
\cite{Courtetal2023} simulate X-ray emission from SNIPs and conclude that the MED time should be larger than $\approx 10^4 \yr$. Other studies suggest much shorter MED times are also possible (e.g., \citealt{Soker2019Rev}).  
\cite{Unoetal2023b} deduce that the CSM mass of SN 2020uem is $0.5-4 M_\odot$, and argue that the CD scenario best explains this SN Ia-CSM (also \citealt{Unoetal2023a}). Other studies of SNe Ia-CSM, i.e., the ejecta interacts with a massive CSM, include \cite{WangLetal2023} and \cite{Yangetal2023} who study SN 2018evt and deduce a CSM mass of $M_{\rm CSM}>0.2M_\odot$. This massive CSM is highly relevant to the present review. \cite{HuMWangetal2023} theoretically study the early excess emission resulting from ejecta-CSM interaction, and \cite{SharmaYetal2023} find that the rate of SNe Ia-CSM is $\approx 0.02\% -0.2\%$ of the SN Ia rate (see also \citealt{Terweletal2024}). Finally, the WD-core merger may lead to an explosion within the envelope, leading to a peculiar type II SN (e.g., \citealt{Kozyrevaetal2024}). 
 
In the double degenerate (DD) scenario (e.g., \citealt{Dimitriadisetal2023, KarthikYadavallietal2023, MoranFraileetal2023, Qietal2023, SrivastavMooreetal2023}) and in the DD-MED scenario, two WDs merge after losing angular momentum to gravitational waves. In the DD scenario, the explosion occurs within a few dynamical timescales, like in the violent merger channel, before the binary system's non-spherical imprints disappear. This DD scenario channel has no MED time, and the explosion is non-spherical. Practically, there are two stars at the explosion, i.e., $N_{\rm exp}=2$. In the DD-MED scenario channel, the explosion occurs only after the merger remnant relaxes. It will likely be a spherical near-$M_{\rm Ch}$ explosion. At explosion, there is only one star (as wide companions play no roles), i.e., $N_{\rm exp}=1$.   

The double detonation (DDet) DDet scenario, where a helium layer accreted from a companion is ignited and detonates the WD, was relatively popular in 2023 (e.g., \citealt{Collinsetal2023, Zingaleetal2023, Boosetal2024}) with attribution of this scenario to some specific peculiar SNe Ia (e.g., \citealt{LiuCMilleretal2023, LiuCetal2023, PadillaGonzalezetal2023ApJ953, PadillaGonzalez2023b} for four peculiar SNe Ia, respectively). 
Several studies attribute hyper-runaway/hypervelocity WDs to the WD helium-donor companion in the DDet scenario (e.g. \citealt{ElBadryetal2023, Werneretal2023}). However, new studies question whether this scenario can explain normal SNe Ia. \cite{Igoshevetal2023}  argue that the numbers and properties of hypervelocity WDs with velocities of $ >1000 \km \s^{-1}$ might at most account for a small fraction of SNe Ia, but might account for some peculiar SNe Ia. \cite{BraudoSoker2023} study the dynamics of the surviving WD and the ejecta, and conclude that very fast runaway WDs require massive exploding CO WD. This makes such a scenario rare and supports the claim that the DDet scenario leads mostly to peculiar SNe Ia but not to normal SNe Ia.

In the single degenerate (SD) scenario (e.g., \citealt{HosseinzadehSandetal2023, LiLiuWang2023, Palicioetal2024}, and \citealt{RuizLapuente2023} and \citealt{ RuizLapuenteetal2023Ap} for Tycho SNR) a WD accretes mass from a non-degenerate companion and explodes when reaching a near-Chandrasekhar mass. It can explode as it reaches the exploding mass, or there might be a long delay from mass accretion to the explosion (MED time). The non-detection of surviving companions in SNRs Ia, e.g., \cite{Shieldsetal2023} for SNR 0509-67.5, \cite{TuckerShappee2023} for SN 2011fe, and \cite{RodriguezYanzaDzib2023} for Tycho's SNR, continues to be a major challenge to the SD scenario (and also to the DDet scenario). 
Considering the difficulties of the SD scenario, I find it unjustified that in 2023 many papers mention only the SD and the DD scenarios, ignoring other SN Ia scenarios.   

The WD-WD collision (WWC) scenario, where the collision of two WDs detonates both of them (e.g., \citealt{Glanzetal2023}),  has been shown over the years to be able to account for no more than a minor fraction of all SNe Ia (see earlier reviews cited in Section \ref{sec:intro}). 

Triple star interaction might enhance the rate of each of these scenarios (e.g., \citealt{Rajamuthukumaretal2023}). Some studies do not support any specific scenario over another (e.g., \citealt{Mengetal2023}). 

Some channels are aimed at peculiar SNe Ia. 
\cite{Zenatietal2023} construct a channel of the DD scenario where the disruption of a low-mass carbon–oxygen (CO) WD by a hybrid helium–CO (HeCO) WD forms calcium-rich peculiar SN Ia and leaves a WD remnant. \cite{Srivastavetal2023} proposed the DD scenario for SN 2022ilv, a super-$M_{\rm Ch}$ peculiar SN Ia. 

Studies in 2023 have continued to explore both sub-Chandrasekhar (sub-$M_{\rm Ch}$) and near-Chandrasekhar ($M_{\rm Ch}$) explosions (e.g., \citealt{Chakrabortyetal2023, DerKacyAshalletal2023xkq, DerKacyetal2023, GuoYWangetal2023, Hoeflich2023, LeungNomoto2023, MaedaJiangetal2023, Nietal2023a, Yarbroughetal2023, Bravoetal2024}). 

The existence of CSM around some SNe Ia continues to be puzzling (e.g. \citealt{Dwarkadasl2023, Maokaietal2023, MoriyaMazzalietal2023, Uchidaetal2024}). However, it is not questionable anymore that some SNe Ia have massive CSM, as \cite{Guestetal2023} find for the SNR 0519–69.0 in the Large Magellanic cloud. Such is SNR G1.9+0.3 that I review in more detail in Section \ref{subsubsec:G1903}. 

The excess emission at early SN Ia phases (within a few days) is still puzzling (e.g., \citealt{LimGetal2023, Nietal2023a, Nietal2023b, WangQResetal2023}). Processes might include the collision of the ejecta with a companion, with a close CSM, or the presence of $^{56}$Ni in the outer zones of the ejecta. Each of these has implications for the possible SN Ia scenarios. \cite{WangQResetal2023} conclude that no current model can adequately explain the full set of observations of the early emission excess of SN 2023bee. It seems that the early light excess cannot yet point to one leading SN Ia scenario but has the potential to do so with more observations and analysis. 

The observation of SNe Ia during the nebular phase (e.g., \citealt{Blondinetal2023, CamachoNevesetal2023, ChenNetal2023, Deckersetal2023, KumarSetl2023, Kwoketal2023, LiuJWangetal2023, Nietal2023a}) is critical to reveal the explosion mechanism as it can, for example, reveal the total mass of some isotopes that are nucleosynthesis at the explosion and their velocity distribution. \cite{ChenNetal2023} and \cite{Kwoketal2023} deduce that SN 2021aefx has a spherically symmetric distribution of observed isotopes at the nebula phase. This supports the notion that many normal SNe Ia have large-scale spherical explosions (for a review, see \citealt{Soker2019Rev}). \citealt{Blondinetal2023} argue that no existing SN Ia explosion model can fit all observed properties of SM 2021aefx.  

The relation between SN Ia types and their environment is another promising direction to learn about SN Ia scenarios, normal and peculiar (e.g., \citealt{Nugentetal2023}).  
Some peculiar SNe Ia might come from a different population than normal SNe Ia. \cite{Barkhudaryan2023} for example, studies the height of SNe Ia and argues that 91T-like peculiar SNe Ia (which are at the slow-declining, hot, luminous end of normal SNe Ia) originate from relatively younger progenitors with ages of $\approx 100 \Myr$, 91bg-like peculiar SNe Ia (which are fast-declining, cool, subluminous peculiar SNe Ia) arise from significantly older progenitors of ages of $\approx ~10 \Gyr$, and normal SNe Ia are from progenitors of $\approx 1 - 10 \Gyr$.

In analyzing observations, one must bear in mind all SN Ia scenarios. This is evident in the study of SN 2020eyj. 
\cite{Kooletal2023} observed SN 2020eyj that they found to be an SN Ia that interacts with a helium-rich CSM. They attribute SN 2020eyj to the SD scenario with a helium non-degenerate donor star. \cite{Kooletal2023} specifically write that their observation ``leave the SD scenario as the only viable alternative''. However, \cite{SokerBear2023} propose a new channel of the CD scenario where there are two major CEE phases. In the second CEE phase, the helium-rich core expands to interact with a WD companion, forming a helium-rich CSM before the explosion. The explosion takes place with a MED time of weeks to tens of years.

\subsection{The group of lonely white dwarf scenarios} 
\label{subsec:LonelyWD}

One of my claims in this review is that many (but not all; see below) of the peculiar SNe Ia result from channels that have two stars at the explosion $N_{\rm exp}=2$, as I present in Table \ref{Tab:Table1}, while normal SNe Ia result mainly from the lonely WD group (e.g., \citealt{BraudoSoker2023}), namely, $N_{\rm exp}=1$. 
I concentrate here on two SNRs that were analyzed in 2023: SN 1181, which is a peculiar SN Iax, and SNR G1.9+0.3, which is probably a normal SN Ia. I first discuss some non-lonely WD scenarios. 

\subsubsection{Peculiar SNe Ia from $N_{\rm exp}=2$ scenarios} 
\label{subsubsec:Nexp2}

I present some recent studies that further motivate me to suggest that most peculiar SNe Ia result from $N_{\rm exp}=2$ scenarios, i.e., scenarios with two stars at the time of the explosion.

From their observation of the super-Chandrasekhar peculiar SN Ia (03fg-like) SN 2020hvf during the nebular phase \cite{SiebertFoleyetal2023} conclude that the explosion is asymmetrical with two velocity components. They discuss the DDet scenario in a double-degenerate system. Namely, one WD exploded, and one survived.  This suggestion, if holds, shows that a scenario with two stars at explosion can lead to an aspherical peculiar SN Ia. That paper also shows the potential of observations during the nebular phase in revealing the exploding system. 

\cite{Kwoketal2023Vmerger} demonstrate the importance of observations during the nebular phase in their study of the peculiar (03fg-like or super-Chandrasekhar) SN 2022pul. The presence of carbon-oxygen rich CSM \citep{SiebertKwoketal2023}, the composition of the ejecta, and the asymmetry of the ejecta led \cite{Kwoketal2023Vmerger} to conclude that the best model of this peculiar SN Ia is the violent merger channel of the DD scenario. This channel has two WDs at the explosion, $N_{\rm exp}=2$. \cite{LiZetal2023} suggest the violent merger channel of the DD scenario for the peculiar SN 2016ije. 
In a different study of `super-Chandrasekhar' SNe Ia, \cite{FitsAxenNugent2023} find the DD scenario to better fit these peculiar SNe Ia than an explosion of a massive magnetized WD. 

The study by  \cite{Hoogendametal2023} is very relevant to the proposition that peculiar SNe Ia have different progenitors than most normal SNe Ia. \cite{Hoogendametal2023} find that 2002es-like and 2003fg-like peculiar SNe Ia have different pre-peak UV color types of evolution compared to normal SNe Ia. They further suggest that these peculiar SNe Ia have different progenitors and propose that these peculiar  SNe Ia may originate in low-metallicity DD or CD scenarios enshrouded by a carbon-rich circumstellar medium. 

\subsubsection{SN 1181 and its SNR Pa 30} 
\label{subsubsec:SN1181}

\begin{figure}[t]
\begin{center}
\includegraphics[trim=2.0cm 18.0cm 6.5cm 0.0cm,scale=0.5]{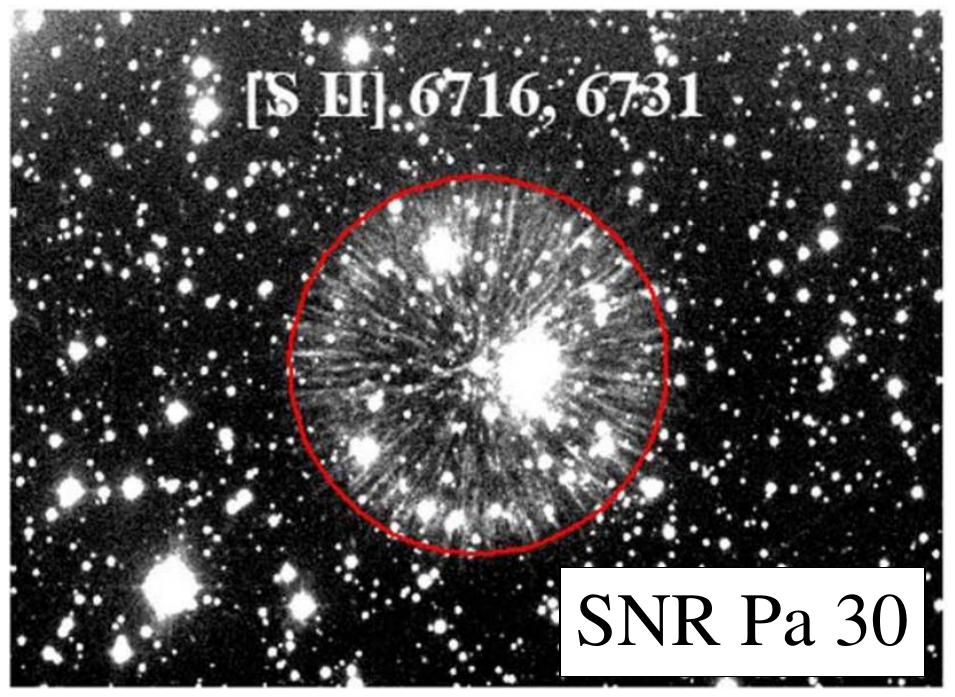} 
\caption{ An [S II] image of SNR Pa 30 from \cite{Fesenetal2023}. This image presents the large-scale spherical morphology of SNR Pa 30, the remnant of SN 1181. It is an SN Iax with a hot remnant at its center. }
\label{Fig:SN1181Pa30}
\end{center}
\end{figure}

Several papers in 2023 revealed the nature of Pa 30 as the SNR of the historical SN 1181, which was an SN Iax \citep{Fesenetal2023, KoTetal2023, Lykouetal2023, Ritteretal2023, Koetal2024}. The large-scale circular morphology of Pa 30 with a hot remnant at its center is critical to the understanding of SN 1181. \cite{Fesenetal2023} conduct a thorough study of Pa 30. They note that the highly spherical morphology of SNR Pa 30 is in sharp contrast to the strongly asymmetric SN Iax ejecta in some models. They attribute the filamentary structure of the round SNR to the photoionization effects of clumpy ejecta (as the modeling of some filamentary planetary nebulae). I refer here only to the large-scale spherical morphology. I, therefore, present one image from \cite{Fesenetal2023} in Figure \ref{Fig:SN1181Pa30}. 

\cite{Schaefer2023} presents a detailed discussion of SN 1181 and its SNR Pa 30. He proposes that the best scenario to match observations is a merger of oxygen-neon (ONe) and carbon-oxygen (CO) WDs (see also \citealt{KoTetal2023}), like in the simulations by \cite{Kashyapetal2018}. However, the simulations by \cite{Kashyapetal2018} show that the outcome is a highly non-spherical ejecta, contrary to the structure of Pa 30. The requirement from scenarios of normal SNe Ia to be able to form spherical SNRs was the highlight of an earlier review of SNe Ia \citep{Soker2019Rev}. This requires the system to relax after a binary interaction, whether a full merger or a mass transfer process. Namely, there must be a merger to explosion delay (MED) time. \cite{MaguireMageeetal2023} study SN 2020udy, which is another SN Iax, and argue that it was an incomplete explosion of a near-$M_{\rm Ch}$ WD. For SN 2020udy, I would also claim for a MED time. The importance of MED time in normal SN Ia scenarios is the highlight of another earlier review \citep{Soker2018Rev}. 

\cite{BogomazovTutukov2023} study the DD scenario with CO WD and one ONe WD for normal SNe Ia (for different outcomes of the merger of a CO WD with an ONe WD, including a possible electron capture CCSN, see \citealt{WuXiongetal2023}). However, SN 1181 is a peculiar SN Ia. This and its spherical ejecta suggest a lonely WD at the explosion time, which resulted from either the CD scenario where either the core of the asymptotic giant branch star or the WD companion, which merged with the core, being ONe rich (a scenario that was studied by \citealt{Canalsetal2018}). Another possibility is the DD-MED scenario where a CO WD and an ONe WD merged to form a near-$M_{\rm Ch}$ WD that exploded after it relaxed, hence the spherical explosion. 

I conclude that SN 1181, with its large-scale spherical SNR, suggests that it results from one of the lonely WD scenarios (Table \ref{Tab:Table1}).  

\subsubsection{SNR G1.9+0.3} 
\label{subsubsec:G1903}

\begin{figure*}[]
	\centering
\includegraphics[trim=1.2cm 8.5cm 2.5cm 2.0cm ,clip, scale=0.5]{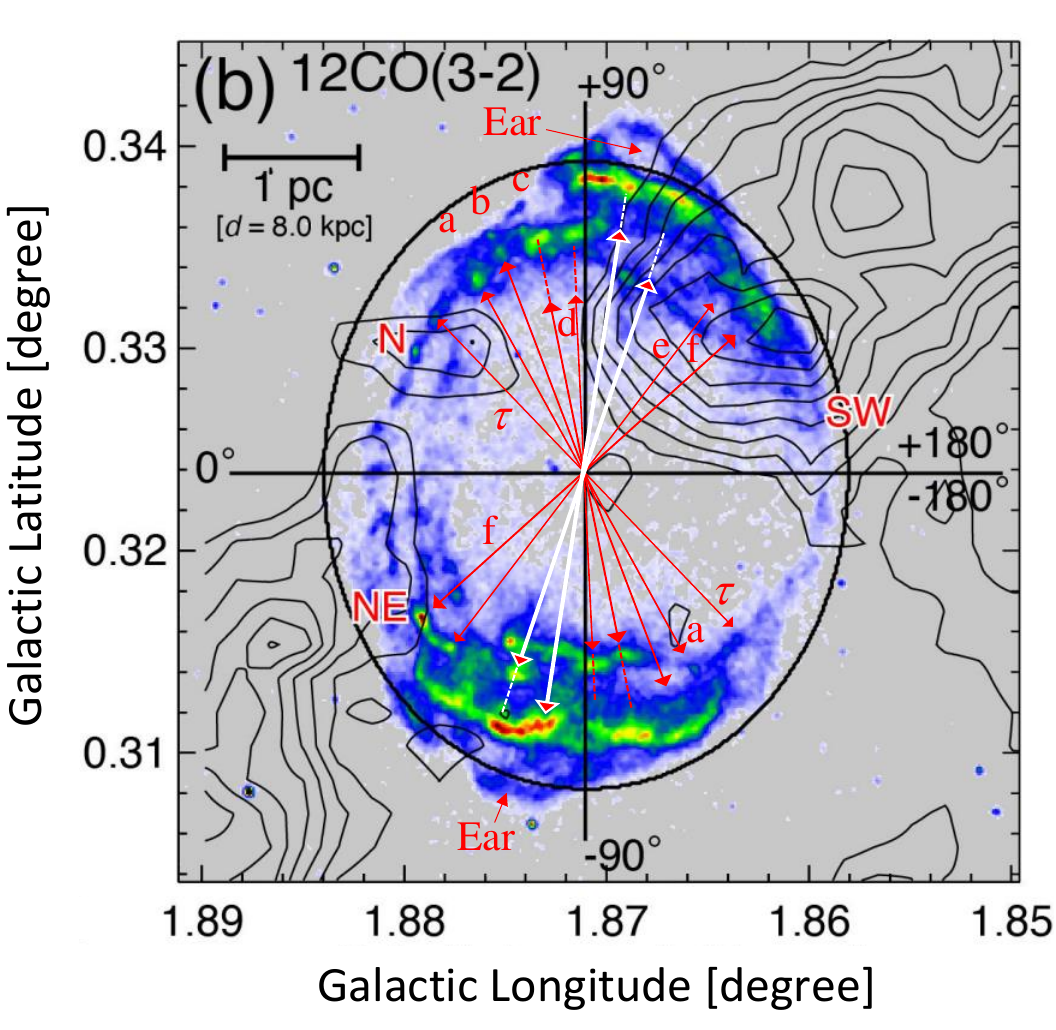}  
\includegraphics[trim=4.03cm 8.5cm 2.5cm 2.0cm ,clip, scale=0.5]{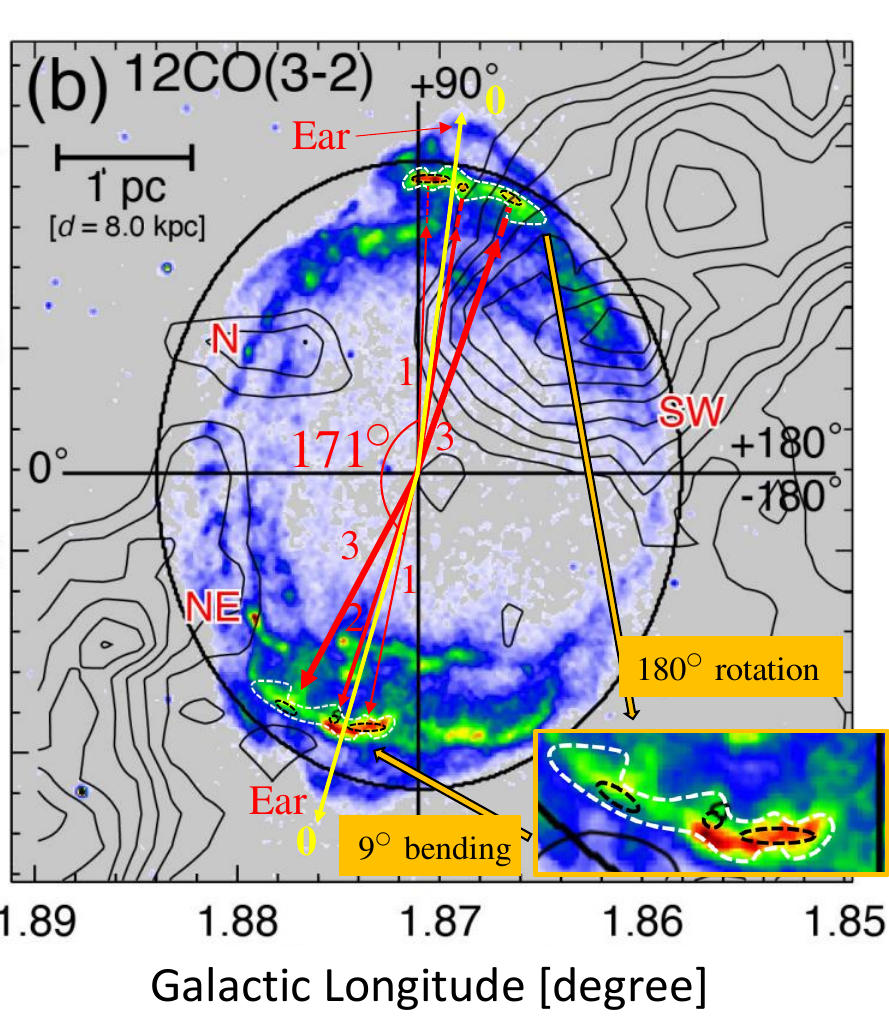}
\caption{The point-symmetric morphology of type Ia SNR G1.9+0.3.
The original X-ray image with contours of CO emission and the notations of the ellipse and coordinates are from \cite{Enokiyaetal2023}.
The double-headed arrows (DHA; red, white, and yellow), the notation of the ears, and the inset on the right panels with the notations of rotation and bending are from \cite{SokerG19032024}. 
The left panel presents the basic point-symmetric morphology. 
Each of the lines DHA-a to DHA-f points at a pair of clumps (twin clumps). The group of twin pairs and the pair of ears form the point-symmetric morphology. Line DHA-$\tau$ is a tentative pair because the clumps are small and faint. The pairs of clumps that two white double-headed arrows point at are treated differently on the right panel. The right panel focuses on the two opposite ears and the arc at the base of each ear. The closed dashed-white line on the upper arc marks the boundary of the arc, and the dashed black lines mark the three emission peaks within the arc. The boundary of the arc with its three peaks is rotated around itself by $180^\circ$ and matched to the arc at the base of the bottom (eastern) ear by a $9^\circ$ displacement as the inset indicates. Namely, there is a $9^\circ$ bend point symmetry of the two ears (DHA-0) and of the arc at the base of each ear (DHA-1 to DHA-3).
}
\label{Fig:G1903}
\end{figure*}

In \cite{SokerG19032024}, I identified a point-symmetry in the morphology of SNR G1.9+0.3, the youngest SNR in the Galaxy, which exploded around 1890-1900. I analyzed the X-ray image of this SNR that \cite{Enokiyaetal2023} released and pointed at ten pairs of opposite clumps or filaments in the X-ray morphology, in addition to the opposite pair of ears that was noticed in the past. I present the analysis in Figure \ref{Fig:G1903} with more details in the caption. Known SN Ia explosion models do not form such point-symmetric structures. I therefore concluded that the point-symmetric structure comes from the CSM which was a planetary nebula. Namely, SNR G1.9+0.3 is an SNIP, i.e., an SN inside a planetary nebula. In \cite{Soker2024IAU384}, I discussed SNR G1.9+0.3 together with the brightest planetary nebulae in the Galaxy and proposed that the planetary nebula into which the progenitor of SNR G1.9+0.3 exploded had a multi-polar morphology. 

To account for the substantial deceleration of the ejecta of SNR G1.9+0.3 \citep{Borkowskietal2017}, the CSM must be relatively massive $M_{\rm CSM} \ga 1 M_\odot$ \citep{SokerG19032024}. Such a massive CSM is more likely to be a planetary nebula that was ejected in a common envelope evolution than the wind from a giant star in the SD scenario. In principle, the DD, DD-MED, and DDet scenarios might each lead to an explosion shortly after a common envelope evolution, namely, before the CSM disperses to the ISM. However, this is very rare because in these scenarios, the core-WD system must lose angular momentum to gravitational waves before the explosion occurs; this is a long process. In addition, SNR G1.9+0.3 was a near-$M_{\rm Ch}$ explosion, as \cite{Kosakowski2023} deduced from the upper bound of $^{44}$Ti abundance that they find. 

I also note that \cite{KobashiLeeetal2023} conclude that the CSM mass of Tycho’s SNR is larger than expected in the SD scenario and better matches the CD scenario. Tycho large-scale circular morphology also suggests a scenario from the lonely WD group of scenarios.

The conclusion of the study of SNR G1.9+0.3 is that it was an explosion inside a planetary nebula of a near-$M_{\rm Ch}$ lonely WD, most likely in the frame of the CD scenario, with a tiny probability for the DD-MED scenario.

\subsubsection{Lonely WDs at the explosion time} 
\label{subsubsec:Lonely}

For its novelty, the group of lonely WD scenarios deserves this subsection. As I indicate in Table \ref{Tab:Table1} the group of lonely WD scenarios includes the DD-MED scenario and the CD scenario. 
The exact definition of a lonely WD explosion is one in which, at the time of the explosion, there is only one star in the system, and it is dynamically relaxed. This implies the explosion might be spherical or axisymmetrical if the lonely WD rapidly rotates. No other material explodes but the WD itself, nor a companion nor an accretion disk.  

Observed properties are more complicated, and not all explosions are spherical. For example, \cite{Nietal2023a} deduce that SN Ia 2021aefx is an $M_{\rm Ch}$ explosion, but the explosion is not spherical. In the lonely-WD scenarios, this is accounted for by an explosion of a rapidly rotating WD or by an off-center deflagration to detonation transition. The rotating WD is likely to sustain a strong magnetic field that might account for the high polarization of some emission lines of the descendant SN Ia. \cite{Yangetal2022} observed high polarization in the Ca II near-infrared triplet of the normal SN Ia 2021rhu, and propose that magnetic field is the cause of this high polarization.   

Consider first the DD-MED scenario. 
The dynamical relaxation timescale of a WD remnant of a merger is about several seconds. An accretion disk around the WD should also be depleted, taking several minutes. Therefore, the MED time should be several minutes or longer. The accretion disk can blow a wind or launch jets, forming a disk-originated matter (DOM). If the MED time is shorter than about a few days, the ejecta collides with the DOM within hours to a few days, leading to an early excess emission (e.g., \citealt{SrivastavMooreetal2023} for SN 2022ywc). Although the DOM might have a bipolar structure, the explosion can be spherical. If the MED time in the DD-MED scenario is much longer, we can have a spherical explosion of a normal SN Ia or a spherical mass ejection in a peculiar SN Ia that leaves a WD remnant after the explosion. The latter channel is a possible scenario for SN 1181 that formed the spherical SNR Pa 30 (Section \ref{subsubsec:SN1181}).  

In the CD scenario, the MED time is longer than several days because the hydrogen-rich envelope that was ejected during the common envelope evolution must be at large distances for the explosion to be classified as SN Ia. In rare cases, the last envelope to be ejected is helium-rich \citep{SokerBear2023}. The CD scenario can explain SNe Ia with massive CSM, up to $M_{\rm CSM} \approx 3 M_\odot$. Since this CSM is or was ionized by the WD merger remnant, it is or was a planetary nebula at the time of the explosion, i.e., the SN Ia is an SNIP. As with many other planetary nebulae that are formed by binary interaction, the CSM might be elliptical or bipolar and possess a point-symmetric morphology. I suggest an SNIP in the CD scenario for SNR G1.9+0.3 (Section \ref{subsubsec:G1903}). The CD scenario with a MED time of $>10^6 \yr$, such that the CSM does not exist anymore, might also account for SN 1181 and its Pa 30 remnant.   

The main theoretical challenge to the solid establishment of the group of lonely WD scenarios is to show that some core-WD and WD-WD merger remnants can live for up to $\approx 10^4 \yr$ and more before they explode (e.g., \citealt{Soker2022CEEDTD}). 
Very encouraging to this challenge is the study by \cite{Neopaneetal2022}, who argue that WD-WD mergers produce a substantial population within a narrow mass range close to the Chandrasekhar mass limit and that the MED time of these remnants might be as long as $\approx 100 \yr$ and more (see also \citealt{IlkovSoker2012}).  \cite{Neopaneetal2022} study the process by which these near-$M_{\rm Ch}$-mass WD merger remnants lose angular momentum and undergo compression and spin-up. These might lead to an explosion. Future studies should further explore this process and explain longer MED times.

\section{Summary} 
\label{sec:Summary}

\subsection{Point symmetric supernova remnants (SNRs)} 
\label{subsec:SNRs2023}

I used the morphologies of CCSNRs and SNRs Ia as a common ground to include both CCSNe and SNe Ia in the same review of SN papers in 2023. 

The research of SNRs is a major topic, with tens of papers in 2023 (Section \ref{sec:PointSymemtry}) that cover many aspects of SNRs. I concentrated mainly on SNR morphologies and, in particular, on point-symmetric morphologies. As I described in Section \ref{sec:PointSymemtry}, the point-symmetric morphology requires that either the explosion itself, as I argue for CCSNe, and/or the CSM, as I argue for SNe Ia, have point-symmetric morphology to begin with. The interaction of a non-point symmetric explosion with a non-point-symmetric CSM or ISM cannot form point-symmetrical SNRs. 

Another common ground for CCSNe and SNRs is that observations during the nebular phase, when the entire ejecta can be observed, might reveal the explosion mechanism and progenitor properties. 

\subsection{Core collapse supernovae (CCSNe) in 2023} 
\label{subsec:CCSNe2023}

Tens of papers in 2023 further demonstrate the diverse properties of CCSNe. Some major topics that research papers of CCSNe in 2023 cover are (Section \ref{subsec:Explosion}) the presence of CSM around CCSNe and the ejecta-CSM interaction,  the role of magnetars in powering superluminous CCSNe, the effects of rapidly rotating pre-collapse cores, and the process of nucleosynthesis in CCSNe. 
SN 2023ixf is a major CCSN of 2023. Many CCSNe have dense CSM at the time of the explosion that is ejected in an outburst shortly before the explosion. The study of SN 2023ixf that had no pre-explosion outburst forces the consideration of a pre-collapse long-lived extended dense zone above the photosphere, like the effervescent zone model (Section \ref{subsec:Explosion}), as an alternative in some cases.

Many papers study the detection of gravitational waves and neutrinos from CCSNe, although these are not yet in a stage to constrain explosion mechanisms. 

In Section \ref{subsec:Breakthrough} I described the analysis of three CCSNRs from 2023 and further discussed the delayed neutrino explosion mechanism and the JJEM.

I summarize the main claims of this review in regard to the explosion mechanism of CCSNe as follows. 
\begin{enumerate}  
    \item The inclusion of energetic magnetars in CCSNe must be accompanied by a discussion of the role of jets because jets accompany the formation of energetic magnetars (Section \ref{subsec:Explosion}).
    \item The delayed neutrino explosion mechanism predicts `failed supernovae', which are not supported by observations (Section \ref{subsec:Breakthrough}). Studies of CCSNe that form black holes cannot ignore jets that, according to the JJEM, explode the star even when the remnant is a black hole. 
    \item According to the JJEM, some CCSNRs might have point-symmetric morphologies. This was verified in three CCSNRs in 2023 (Figures \ref{Fig:N63A} - \ref{Fig:SN1987A}; for a discussion of the total 8 CCSNRs with identified point-symmetric morphologies, see \citealt{Soker2024CFPointSymmetry}).  The main caveat in the identification of eight-point-symmetric CCSNRs is the lack of quantitative classification criteria. This is the subject of ongoing research.
    I consider this group of three-point-symmetric CCSNRs (and a speculative identification of a point-symmetric morphology in SNR CTB~1; \citealt{BearSoker2023}) to constitute a 2023 breakthrough in our quest for the explosion mechanism of CCSNe, which, I suggest, is the JJEM. 
\end{enumerate}

\subsection{Type Ia supernovae (SNe Ia) in 2023} 
\label{subsec:SneIa2023}

In Section \ref{sec:SNeIa} I reviewed the scenarios of normal and peculiar SNe Ia that have been studied in 2023. I presented the classification into six scenarios in Table \ref{Tab:Table1} and criticized studies that, even in 2023, still mention only the SD and DD scenarios and ignore other scenarios, particularly the CD and DD-MED scenarios.   

The main conclusions related to SNe Ia are as follows. 
\begin{enumerate}
     \item The diversity of both normal and peculiar SNe Ia suggests that more than just two scenarios are involved in forming most of all types of normal and peculiar SNe Ia (likely 4 to 5 scenarios). Studies should be open to the different scenarios (Table \ref{Tab:Table1}). The habit of mentioning only the SD and DD scenarios should be abandoned. 
    \item Studies in 2023 have shown that observations during the nebular phase could deduce that some peculiar SNe Ia result from some specific scenarios, mainly, but not only, from non-lonely scenarios, i.e., $N_{\rm exp}=2$ (Sections \ref{subsec:SNIaScenarios} and \ref{subsubsec:Nexp2}).  
    \item  As noticed by past reviews of SNe Ia, the large-scale spherical morphologies of many SN Ia remnants constrain the common SN Ia scenario(s) to be able to avoid aspherical effects, such as a close companion. The large-scale spherical structure of Pa 30, the remnant of the peculiar SN Ia 1181 (Section \ref{subsubsec:SN1181}), and other considerations in 2023 have motivated the introduction of the group of lonely WD scenarios (Table \ref{Tab:Table1}).   
    \item  The point symmetric morphology of SNR G1.9+0.3 is the strongest ever argument for SN Ia inside a planetary nebula (SNIP; Section \ref{subsubsec:G1903}). Together with SNR Pa 30 of SN 1181, they also put the SN Ia lonely WD scenarios, which have MED time, on very solid ground.   
    \item Taking the above conclusions, the group of lonely WD scenarios (and non-lonely WD scenarios) should be a major lead in our quest to related SN Ia scenarios (Table \ref{Tab:Table1}) with observed normal and peculiar SNe Ia.    
\end{enumerate}

More research, like the search and detection of supernova progenitors, both CCSNe (e.g., \citealt{Strotjohannetal2023, VanDyketal2023}) and SNe Ia, are essential to verify or reject the breakthroughs I pointed out in this review. Essential to the study of explosion mechanisms and progenitor properties is to consider both theoretical explosion mechanisms of CCSNe and all SN Ia scenarios.

\section*{Acknowledgements}
I thank Hila Glanz, Amit Kashi, and two anonymous referees for the useful comments. 
In writing this review I was motivated by the results and discussions of two 2023 meetings on supernovae, \textit{SuperNova EXplosions Conference} (SNEX, Technion, Israel; chaired by Hagai Perets and Yossef Zenati) and \textit{SuperVirtual 2023} (virtual; chaired by Melina Cecilia Bersten and Takashi Moriya). 
This research was supported by grants from the Pazy Research Foundation and a grant from the Israel Science Foundation (769/20).


\label{lastpage}
\end{document}